\documentclass[letterpaper]{article} % DO NOT CHANGE THIS
\usepackage{aaai2026}  % submission option removed to show authors; revisit for final copyright per camera-ready instructions
\usepackage{times}  % DO NOT CHANGE THIS
\usepackage{helvet}  % DO NOT CHANGE THIS
\usepackage{courier}  % DO NOT CHANGE THIS
\usepackage[hyphens]{url}  % DO NOT CHANGE THIS
\usepackage{kotex} % Added for Korean character support
\usepackage{graphicx} % DO NOT CHANGE THIS
\usepackage{natbib}  % DO NOT CHANGE THIS AND DO NOT ADD ANY OPTIONS TO IT
\usepackage{caption} % DO NOT CHANGE THIS AND DO NOT ADD ANY OPTIONS TO IT
\usepackage{subcaption}
\usepackage{multirow}
\usepackage{amsmath}  % needed for \text{} inside math mode
\usepackage{algorithm}
\usepackage{algorithmic}

\usepackage{booktabs}

\usepackage{newfloat}
\usepackage{listings}
\usepackage{pdfpages}  % appends the supplement as Appendices A-I
\DeclareCaptionStyle{ruled}{labelfont=normalfont,labelsep=colon,strut=off} % DO NOT CHANGE THIS
\floatstyle{ruled}
\newfloat{listing}{tb}{lst}{}
\floatname{listing}{Listing}
\title{When AI Becomes Routine: \\A Decade of Public AI Mediation in Korean Go Commentary}
\author{
    Haewoon Kwak
}
\affiliations{
    Indiana University Bloomington\\
    hwkwak@iu.edu
}

\begin{document}

\maketitle

\begin{abstract}
When AI systems surpass elite human performance and settle into everyday expert practice, the question that follows is how machine judgment is made publicly intelligible and attributable. We study Korean Go commentary on YouTube, where AI systems such as KataGo became standard analytic tools after AlphaGo. Our corpus spans a decade (2016--2025) and approximately $1{,}900$ hours of footage across institutional broadcasters and creator-led channels, in four phases of AI availability. We document a widening asymmetry between visual and verbal AI presence: AI winrate graphs are visible for about $98\%$ of late-period institutional broadcast time, yet AI-salient talk accounts for only $2.63\%$ of sentences. What recedes is the source label, not the metric: winrate and point-gap talk persists while ``AI'' itself goes unsaid. We read this recession as the communicative signature of domestication. Our strongest evidence is a compositional shift in verbal mediation: explicit naming gives way to interface rendering, and creator-led commentary leans further toward it than institutional commentary. We develop a typology distinguishing source-foregrounding from source-receding mediation, and argue that the two preserve different hooks of contestability: discursive anchors through which audiences can recognize and question the machine source. The stakes of that difference rise in domains where AI is less reliable than in Go.
\end{abstract}

% This block must stay between (not within) the abstract and the main body.
% Add \link{Extended version}{...} here once the arXiv ID is announced.
\begin{links}
    \link{Code}{https://github.com/haewoon/routine-superhuman-ai}
\end{links}

\section{Introduction}

As AI systems that surpass top human performance become embedded in everyday work, the question is how their outputs are made intelligible, attributable, and contestable in routine public settings. In many such settings, AI judgment becomes public not because the model explains itself, but because human intermediaries learn how to speak with, around, and sometimes against machine outputs. Existing debates on AI and expertise often focus on anticipation, attitudes, or short-run reactions \cite{brauner2025mapping,miyazaki2024public}. We know less about what happens after a community has lived for years with routine access to superhuman systems.

\looseness=-1 The game of Go offers a rare case of such a mature setting. Played on a 19$\times$19 grid, Go has no piece-by-piece material count: position value depends on holistic territorial estimation across competing frameworks whose boundaries resolve only late in the game, and the branching factor is roughly an order of magnitude above Chess. Positional judgment in the opening and middle game is an intuitive task that even strong professionals cannot exhaustively calculate, especially under live time pressure. AlphaGo's 2016 victory over Lee Sedol \cite{silver2016mastering} was therefore experienced as a substantive break rather than a faster-search milestone. Open-source AI systems such as KataGo \cite{wu2019accelerating} subsequently made that capability a routine analytic resource, making visible a winrate and recommended move for any position in seconds where a professional review session might previously have taken hours. We use \textit{routine superhuman AI} to refer to a social condition in which AI systems that surpass elite human performance are routinely available inside everyday expert practice. Korean Go is especially valuable because AI did not remain an abstract benchmark: documentary, pedagogical, and broadcaster-facing materials now describe AI-assisted review rooms, routine winrate overlays, and commentary practices organized around interpreting AI output for viewers \cite{ebs2026badukdoc,monthlybaduk2021ai,kim2024opening70,baduktv2020format}.

\looseness=-1 Our empirical focus is not the AI systems' playing strength itself, but the public mediation of machine judgment in commentary. Commentary is where expertise is narrated in real time for an audience, and where AI outputs can be named, translated, softened, or resisted. This mediation layer is governance-relevant: machine judgment becomes publicly intelligible, attributable, and contestable through it. We study Korean Go commentary on YouTube, which functions both as the primary platform for creators and as the public archive for institutional broadcasters. We assemble a large-scale corpus spanning a decade (2016--2025): $609$ unique source videos and approximately $1{,}900$ hours of footage. This shared platform makes it possible to compare institutional and creator-led commentary while holding distribution infrastructure relatively constant. Because only \textit{BadukTV} provides consistent archival coverage across all four phases, the longitudinal backbone is necessarily BadukTV-centered, while the creator comparison is strongest in the late-routine period.

From this framing, two research questions follow:

\begin{itemize}
    \item RQ1: How does overt verbal AI activation change over time in the institutional longitudinal backbone of Korean Go commentary? (\S\ref{sec:longitudinal})
    \item RQ2: In the late-routine period, how do institutional and creator-led channels differ in the amount and composition of overt AI-salient commentary? (\S\ref{sec:channel_comparison})
\end{itemize}

We also report a within-speaker follow-up test using commentator Lee Hyunwook across institutional and personal media settings (\S\ref{sec:case_study}), and a small exploratory audience extension in live videos with archived chat (\S\ref{sec:audience_uptake}). Both serve as bounded extensions rather than standalone research questions, with the commentator's mediation layer remaining the paper's core empirical backbone.

\looseness=-1 We make three contributions. First, we introduce a conservative keyword-anchored \textit{AI-salient subset} that treats explicit AI naming as a marked surface form (a residual moment of overt source attribution against an already-embedded AI background), and check its sensitivity with alternative per-word and per-minute denominators. Second, we develop a conceptual account of public AI mediation: the communicative layer in which experts name, translate, soften, or resist machine judgment for audiences. From this account, we derive a mediation-form typology distinguishing source-foregrounding from source-receding practices, grounded in a qualitative inspection of recurring micro-patterns (\S\ref{sec:micro_patterns}). Each form preserves different hooks of contestability, and the retreat of explicit naming becomes the signature of domestication. Third, we show how this layer matters for AIES because accountability and contestability depend not only on what a model outputs, but also on how that output is made publicly intelligible and attributable after deployment. The paper's strongest evidence is a compositional shift in verbal mediation, from explicit naming toward interface rendering, not legitimacy or governance outcomes directly. Whether that shift reflects a deeper change in how experts treat AI judgment or a broadcast convention against narrating an on-screen graphic is underdetermined by verbal data alone, a limit we carry through the discussion.

\section{Background and Related Work}

\paragraph{Why Go is a useful test case.}
Elite Go play was long described in terms of intuition rather than calculation \cite{gelly2012grand}, and AlphaGo's break with that history disrupted not only rankings but a public model of professional judgment \cite{silver2016mastering}. Unlike Chess, where reflection on machine superiority centered on earlier milestones such as Deep Blue \cite{ensmenger2012chess}, Go experienced a sharper break between human-centered expertise and routine, widely accessible machine superiority, making it a clean case for observing both initial shock and later domestication.

\paragraph{AI in Professional and Creative Domains.}
\looseness=-1 The transformation of professional expertise by AI spans high-stakes and creative domains: in medicine, AI diagnostics create friction between clinical intuition and algorithmic probability \cite{lebovitz2022engage}; in creative work, generative systems challenge norms of authorship and evaluative authority \cite{epstein2020gets}. The behavioral literature documents conflicting tendencies: automation bias \cite{skitka1999does}, algorithm aversion \cite{dietvorst2015algorithm}, algorithm appreciation \cite{logg2019algorithm}, and trust dynamics that vary across time and occupations \cite{glikson2020human,miyazaki2024public}. We instead study what comes after adoption. Once high-performing AI is already routine professional infrastructure, expert commentary repeatedly mediates AI outputs for a public audience in a naturalistic setting. This connects to AIES concerns after deployment: the issue is how outputs become interpretable, contestable, and usable in public settings \cite{doshi2017towards,miller2019explanation,rahwan2018society,shneiderman2020hcai}.

\paragraph{Explanation, Contestability, and Mediatization.}
\looseness=-1 Recent AIES and HCI work emphasizes that explainability is not exhausted by model internals or standalone interfaces. Studies of explainability in practice show that users ask situated questions, rely on organizational context, and often struggle to use interpretability tools as designers expect \cite{liao2020questioning,kaur2020interpreting,ehsan2021expanding}. Explanation is partly accomplished in use: through the people, routines, and communicative formats that render output meaningful for others. This resonates with gatekeeping research treating intermediaries as selectors who determine what information reaches publics \cite{shoemaker2009gatekeeping}. We build on both by treating commentators as \textit{public interpreters of AI output}: intermediaries who not only consume model output but perform explanation in real time.

This same literature highlights accountability and contestability as governance concerns: contestable-AI work argues that affected publics need ways to question and obtain review of AI-mediated judgments \cite{almada2019human,lyons2021conceptualising,alfrink2023contestable,castleman2024why}, and accountability depends on documenting how AI outputs are translated into decisions and institutional action \cite{raji2020closing,metcalf2021impact}. Some mediation forms foreground the machine source while others fold it into familiar metrics. This distinction connects to contestability: explicit naming preserves discursive anchors for challenge while interface rendering may erode them.

Mediatization theory further suggests that media logics (the norms, formats, and routines of a given medium) shape how expertise itself is performed \cite{hjarvard2008mediatization}. In our setting, institutional television simulcasts and creator-led streaming impose different production constraints on the same underlying AI infrastructure.

\paragraph{Authority, Expertise, and Domestication.}
\looseness=-1 Domestication theory \cite{silverstone1996design,berker2006domestication,hartmann2013domestication} describes how users tame and integrate new technologies into existing practices. The more recent strand emphasizes that domestication is an ongoing process in which the technology becomes part of how everyday authority is produced. We use this framework to understand how the Go community transformed AlphaGo's initially alien logic into standard practice, and how the technology's recession into the unmarked background of expert speech is itself a key signature of that process. Affordance theory \cite{gibson1979ecological, hutchby2001technologies} further suggests that the two channel types in our data operate under different production logics: institutional broadcasters simulcast TV programming with fixed schedules, multi-commentator formats, and editorial oversight, while creators leverage platform affordances such as real-time chat and solo streaming. Korean reportive forms (quotative endings that relay another's statement) add a culturally specific resource for relaying machine judgment that is grammatically indirect but pragmatically authoritative, making Korean commentary a setting where cultural-linguistic affordances themselves shape the mediation repertoire.

\section{Data and Methods}

To analyze shifts in public-facing mediation, we constructed a longitudinal dataset spanning ten years (2016--2025), covering the period before and after AlphaGo. We divide the timeline into four analytic phases anchored in observable changes in the availability and routinization of AI analysis:

\begin{itemize}
    \item Phase 1: Human Mastery (Jan 2016 -- Feb 2016): the pre-AlphaGo baseline, before AI stronger than human professionals entered everyday commentary.
    \item Phase 2: Shock (Mar 2016 -- Dec 2017): the immediate aftermath of AlphaGo vs. Lee Sedol, when commentators publicly grappled with machine superiority.
    \item Phase 3: Diffusion (Jan 2018 -- Dec 2020): AI analysis became increasingly available, especially through open-source AI systems and creator experimentation.
    \item Phase 4: Routine Integration (Jan 2021 -- Dec 2025): AI analysis appears as stable infrastructure in commentary rather than an exceptional novelty.
\end{itemize}

We place the Phase~3/4 boundary at January~2021, not as a sudden industry breakpoint but as the first full window in which AI-mediated commentary appears across both institutional and creator-led channels as an expected component of live explanation, following the consolidation of open-source AI systems such as KataGo as standard analytic tools. Appendix~D reports the same comparisons under five boundaries from January~2020 to January~2024.

\paragraph{Data Sources and Curation.}
We selected four Korean-language channels to capture variation in media context while keeping domain and audience comparable, applying inclusion criteria for sustained Go-commentary activity, clear expert anchoring, archive depth, and public reach. The final sample covers two institutional broadcasters, \textit{BadukTV} \cite{baduktvchannel} and \textit{K-Baduk} \cite{kbadukchannel} (Korea's two major cable Go channels simulcasting to YouTube), and two creator-led professional-player channels, \textit{ProYeonwoo} \cite{proyeonwoochannel} (Cho Yeonwoo, Pro 2-dan, active since Jun 2015) and \textit{LeeHyunWookTV} \cite{leehyunwooktvchannel} (Lee Hyunwook, Pro 9-dan, launched Sep 2019). Among the institutional channels, only \textit{BadukTV} provides consistent four-phase coverage. Appendix~A details subscriber counts and channel-by-channel programming styles.

\begin{figure}[t]
\centering
\includegraphics[width=\columnwidth]{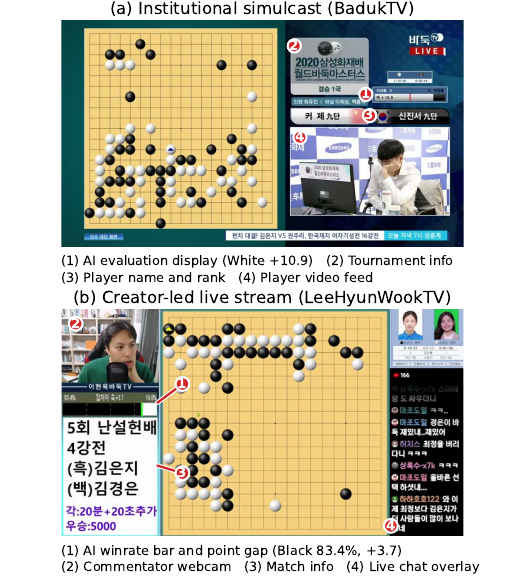}
\caption{Typical broadcast layouts. (a) Institutional simulcast (\textit{BadukTV}, 2020 Samsung Cup final). (b) Creator-led live stream (\textit{LeeHyunWookTV}, 2024 Nanseolheon Cup). In both, the AI evaluation layer is on screen whether or not commentators verbally name it.}
\label{fig:broadcast_layouts}
\end{figure}

Figure~\ref{fig:broadcast_layouts} shows typical on-screen layouts for the two channel types. In both, AI-derived evaluations are continuously visible beside the board: an evaluation display in the institutional simulcast, a winrate bar and point-gap estimate in the creator stream. Creator streams add platform affordances such as webcam narration and live chat. This continuously visible AI layer is the background against which our verbal measures operate.

\paragraph{Video Selection Strategy.}
We constructed two core datasets ($D_{Long}$ and $D_{Case}$) and one follow-up dataset ($D_{Creator}$). Table~\ref{tab:dataset_overview} summarizes the design of each dataset. Full per-dataset selection criteria are in Appendix~A.

\begin{table*}[t]
\centering
\small
\begin{tabular*}{\textwidth}{@{\extracolsep{\fill}}llp{4.0cm}p{8.1cm}r@{}}
\toprule
Group & Dataset & Channels & Design & Videos \\
\midrule
\multirow{2}{*}{Core} & $D_{Long}$ & BadukTV, K-Baduk, LeeHyunWookTV, ProYeonwoo & Four-phase backbone; 10 channel-phase strata, targeted at 40 videos each. Creator channels appear in Phases 3--4 only. & 400 \\
& $D_{Case}$ & BadukTV, LeeHyunWookTV & Lee Hyunwook only. Matched institutional/personal corpus (57+57) across Phases 3--4. & 114 \\
\midrule
Follow-up & $D_{Creator}$ & ProYeonwoo, LeeHyunWookTV, ChoHyeyeon, RyuSihunWorld, DongneBaduk & Phase 4 only. Five creator channels, balanced across 2023 and 2024 (10 videos per year within channel). & 100 \\
\bottomrule
\end{tabular*}
\caption{Overview of the three datasets. Counts are dataset entries. The follow-up dataset overlaps slightly with the core corpora.}
\label{tab:dataset_overview}
\end{table*}

\looseness=-1 $D_{Long}$ is the longitudinal backbone: a stratified sample of $400$ high-visibility videos ($\sim 1{,}394$ hours) drawn from more than $28{,}000$ candidates, spanning the four phases and supporting RQ1. The sample is built around live broadcasts of major professional matches (especially international tournaments) and post-game reviews. Our estimand is the commentary in these videos, where AI-mediated expert judgment is most likely to matter, not the average Korean Go video. The selected videos have substantially higher average view counts than unselected candidates ($80{,}147$ vs.\ $10{,}596$; $p < 0.001$). $D_{Case}$ supports the within-speaker follow-up test (\S\ref{sec:case_study}): $57$ \textit{BadukTV} television appearances by Lee Hyunwook paired with $57$ temporally proximate live-game streams from his personal channel.

A follow-up dataset, $D_{Creator}$, extends the Phase~4 creator analysis to five channels balanced across 2023 and 2024, adding \textit{ChoHyeyeon} \cite{chohyeyeonchannel}, \textit{RyuSihunWorld} \cite{ryusihunworldchannel}, and \textit{DongneBaduk} \cite{dongnebadukchannel} to the two creator channels above, and supporting the broader field-level contrast in \S\ref{sec:channel_comparison}. Across the three datasets the study uses $614$ dataset entries corresponding to $609$ unique source videos. Appendix~A maps each empirical claim onto the datasets it draws on. The AI-salient analysis denominators are $D_{Long}$ ($400$ listed / $398$ transcribed videos / $673{,}557$ sentences after non-game filtering) and $D_{Creator}$ ($100$ listed / $100$ transcribed / $116{,}202$ sentences). The exploratory topic analysis below uses SBERT and a BERTopic-style pipeline on the larger pre-filter main corpus ($876{,}353$ sentences across $D_{Long}$ and $D_{Case}$), because that pipeline does not exclude non-game segments; we use those topics only as scaffolding rather than as the backbone of quantitative claims.

\paragraph{Transcription and Sentence Segmentation.}
Transcripts were generated with Whisper Base \cite{radford2023whisper} and reconstructed into sentences with the Kiwi Korean morphological analyzer \cite{bab2min2024kiwi}. We selected Whisper Base because (i) our use case is keyword extraction rather than verbatim transcription, and (ii) a targeted 120-row STT keyword audit found acceptable transcription in 98.3\% of items and 100\% preservation of AI-anchor terms (the inputs to our keyword pipeline). Because sentence segmentation may vary with speech style, all main volume claims are also checked against per-word and per-minute rates.

All quoted examples are translated by the author, a native Korean speaker, and appear in the body in English translation, with Korean originals in Appendix~I. Translations aim for pragmatic equivalence rather than word-for-word correspondence. Quoted Korean is lightly normalized from raw STT output for readability, with normalization limited to unambiguous transcription errors.

\paragraph{Coding and Reliability.}
All primary coding (precision audits, micro-pattern identification, audience chat categorization, and false-negative spot-checks) was performed by the author using fixed codebooks. We treat this single-coder primary design as a methodological constraint and bound what each reliability check below can support.

We assessed coding reliability through two complementary checks on a stratified 100-sentence sample (95 effective for intra-rater, 94 for inter-rater after excluding STT-degraded or unclear cases). Intra-rater test--retest (lead coder, blind, two weeks apart) yielded Cohen's $\kappa = 0.98$ for the binary AI-salient classification (95\% CI [0.94, 1.00]) and marker-level $\kappa$ between 0.87 and 1.00. External agreement with a second coder (native Korean, not Go-familiar) yielded $\kappa = 0.98$ binary and marker-level $\kappa$ between 0.42 and 0.89. The lower marker-level values concentrate on a single Go-context-dependent marker, which we report descriptively only. Load-bearing inferential claims do not depend on it. Bootstrap stability checks on the precision audit (stratified 95\% CI [0.93, 0.99]) and false-negative audit (miss-rate 95\% CI [0.000, 0.027]) of the AI-salient subset are reported in Appendix~E, along with per-marker $\kappa$ values, the per-item diagnostic (Appendix~F), and the codebook.

\paragraph{Analytic Scope and Statistical Reporting.}
Our main empirical backbone is commentator discourse. A targeted OCR-based visibility check on 54 Phase~4 institutional broadcasts using per-video manually-annotated winrate-graph regions of interest (calibration procedure in Appendix~C) found AI winrate graphs visible for an average of 97.99\% of broadcast time (median 99.83\%; 51 of 54 videos $\geq$95\% visible). Because Phase~3 uses a different AI graphic format, we built a Phase~3-specific detector and applied it to a separate \textit{BadukTV} sample ($N=6$). Under this phase-specific detector, AI-graphic visibility in the Phase~3 \textit{BadukTV} sample was a mean of 86.9\% of broadcast time (median 88.0\%; range 81.7--92.8\%). The visual AI layer was therefore already substantially embedded in Phase~3, and the Phase~3 to Phase~4 verbal compositional shift toward interface-only mediation cannot be straightforwardly explained by the visual layer first appearing in Phase~4. The transcript-based measure thus captures verbal uptake of an already-embedded visual AI layer in the institutional sample, not the presence of visualization as such.

We use video-level Welch and paired $t$-tests as descriptive summaries of planned contrasts and replicate the main volume comparisons with two alternative denominators (AI-salient sentences per 1,000 words and per minute) as unitization sensitivity checks. We also verified that the main patterns survive alternative Phase~3/4 boundary shifts at January~2022 and January~2024.

\section{AI-Salient Analysis Design}
\label{sec:ai_salient_design}
Our empirical entry point is not a sentence-wide taxonomy, but the subset of commentary moments in which AI becomes overtly verbally activated. Early exploratory reading and lexical comparison (Appendix~B) suggested that the post-AlphaGo period was characterized less by a total replacement of human explanation than by new public vocabularies for rendering machine judgment. The \textit{marked surface form} framing of overt AI naming is motivated by domestication of post-routinization technology talk \cite{silverstone1996design,berker2006domestication}. We operationalize this through a precision-first keyword-anchored \textit{AI-salient subset}, the paper's main backbone, which anchors measurement directly in observable text rather than relying on a sentence-wide classifier.

\subsection{Keyword-Anchored AI-Salient Subset}
An \textit{AI-salient} sentence contains either: (1) an explicit AI-reference marker such as \textit{AI}, \textit{인공지능} (AI in Korean), \textit{카타고/KataGo}, or \textit{절예/Jueyi}; or (2) interface/metric language through which AI output is publicly rendered and discussed, such as \textit{graph}, \textit{블루스팟/bluespot} (the AI's top recommended move, displayed in blue on the board), winrate terms, bounded percentage expressions, or decimal point-gap estimates such as \textit{1.6집} (1.6 points). Because bare \textit{승률} (winrate) and percentage language can also describe player records, rankings, or other non-AI contexts, the strict rule excludes obvious record/performance and advertisement contexts and treats broader point-gap language such as \textit{집차} as AI-salient only when it is locally anchored by adjacent AI/interface cues. This design prioritizes precision over recall. Table~\ref{tab:keyword_inventory} summarizes the rule families used in the paper.

Within the AI-salient subset, we track four markers:
\begin{itemize}
    \item \textbf{Explicit AI naming}: direct mention of AI, KataGo, or another AI system.
    \item \textbf{Interface / metric}: graph, winrate, point-gap, bluespot, or comparable interface language.
    \item \textbf{Frictional uptake}: co-occurring lexical markers of difficulty, disbelief, or interpretive strain.
    \item \textbf{Indirect citation}: reportive phrasings such as ``it says ...'' or ``it is saying ...'' that relay machine judgment.
\end{itemize}

These indicators are not mutually exclusive. Because subset entry already depends on explicit or interface-style anchors, we do not interpret the internal composition as a single ``winner'' category. Instead, we report overlap patterns between explicit naming and interface/metric mediation to distinguish explicit-only, interface-only, and jointly anchored cases inside the subset.

We also construct an \textit{expanded} sensitivity version where reportive forms in game-progress contexts can anchor a sentence on their own, without an adjacent AI/interface marker. A 120-sentence audit found 116/120 precision (96.7\%). The expansion changes overall AI-salient shares only marginally (e.g., \textit{BadukTV} Phase~4: 2.63\% to 2.67\%), so we use the stricter version as the main quantitative backbone, and describe reportive relay qualitatively in the micro-pattern section below.

A stratified manual audit of 150 strict-backbone hits found 145/150 true positives (96.7\%, 95\% Wilson CI [92.4, 98.6]). A complete audit of all 13 strict hits in Phase~1 found them all to be false positives from generic percentage talk. We treat the pre-AlphaGo baseline as effectively zero. A bounded false-negative spot-check on 255 sentences from ten late-period windows found only three additional overt AI misses beyond the three already captured: two in context-dependent metric talk, one in a Korean Go community wordplay that invokes AI through a player nickname (\textit{신공지능}, fusing top player Shin Jin-seo's name, \textit{신진서}, with \textit{인공지능}). The strict backbone is therefore a high-precision conservative lower bound.

\paragraph{What the Measure Captures}
\looseness=-1 This backbone does not measure total AI dependence, and we deliberately resist treating it as such. In Phases~3--4, winrate graphs and AI-recommended moves are typically already on screen and routinely shape commentator reasoning even when AI is not named explicitly. Against the 97.99\% winrate visibility reported above, AI is verbally activated in only 2.63\% of \textit{BadukTV} Phase~4 sentences, so ``how much AI is being used'' is no longer a single quantity a verbal subset could approximate. We instead treat explicit AI naming as a \textit{marked} surface form: a moment in which commentators name or quote a source that has otherwise become the unmarked evaluative ground of late-routine commentary (i.e., Phase~4). Read this way, the strict subset measures the residual marked moments of overt source attribution, and its declining share relative to interface-only mediation is itself the communicative signature of domestication \cite{silverstone1996design,berker2006domestication,hartmann2013domestication}. The analytically interesting question is therefore not how small the share is, but what pragmatic functions explicit naming continues to perform once AI has receded into the unmarked default. Qualitative inspection (\S\ref{sec:micro_patterns} below) suggests recommendation citation, delegated voice through Korean reportive forms, and human--machine contrast as recurring contexts.

\looseness=-1 A behavioral alternative to this reading is worth addressing: commentators might name AI less simply because the AI-derived metric is permanently visible on screen, much as an on-screen scoreboard reduces verbal score-citation. Our data point against this simpler suppression mechanism. Winrate readings and point-gap estimates (visible on screen at near-saturation) continue to appear verbally throughout late-routine commentary. What disappears is the source label ``AI'' itself: this is naturalization of the source attribution, not suppression of the metric talk.

\subsection{Exploratory Audience Uptake Around AI-Salient Commentary Events}
We constructed an event-linked audience sample from videos with archived live-chat replay, yielding 48 event--baseline pairs and 619 hand-coded chat messages. We treat this audience layer as a targeted extension rather than a second backbone (details in Appendix~H).

\begin{table*}[t]
\centering
\footnotesize
\begin{tabular*}{\textwidth}{@{\extracolsep{\fill}}p{2.3cm}p{5.0cm}p{2.0cm}p{5.5cm}@{}}
\toprule
\textbf{Keyword family} & \textbf{Representative forms} & \textbf{Rule tier} & \textbf{Analytic role} \\
\midrule
Direct AI reference & \textit{AI}, \textit{인공지능}, \textit{카타고/KataGo}, \textit{절예/Jueyi} & Strict & Marks overt naming of the machine source. Other candidate tokens (\textit{엔진} `engine', \textit{릴라/Leela Zero}, bare \textit{모델/프로그램/시스템} `model/program/system') were inspected and excluded as too rare, too generic, or non-AI in this corpus. \\
Interface / metric markers & \textit{graph}, \textit{블루스팟/bluespot}, \textit{winrate}, \textit{승률}, \textit{2.9\%}, \textit{1.6집} & Strict & Captures publicly legible interface and metric cues through which machine judgment is rendered usable. \\
Context-anchored point-gap & \textit{집차} only when locally anchored by adjacent AI/interface cues & Strict & Prevents ordinary human count talk from being overcounted as AI-salient. \\
Reportive indirect citation & ``$\sim$한다고 합니다'', ``$\sim$라고 하네요'' with metric or move-recommendation context & Expanded sensitivity & Captures routinized relay of machine judgment without repeated overt naming. \\
Friction / translation candidates & ``for a human ...'', ``what this means is ...'', ``it says this is bad'' & Exploratory only & Guides qualitative inspection of recurrent micro-patterns but is not used as a backbone rule family. \\
\bottomrule
\end{tabular*}
\caption{Keyword inventory used in the AI-salient analysis.}
\label{tab:keyword_inventory}
\end{table*}

% -------------------------------------------------------------------------
\section{Results}

\begin{table*}[t]
\centering
\footnotesize
\begin{tabular*}{\textwidth}{@{\extracolsep{\fill}}lp{2.4cm}p{2.2cm}p{2.2cm}p{2.2cm}p{2.2cm}@{}}
\toprule
\textbf{Setting} & \textbf{AI-salient share} & \textbf{Explicit AI naming} & \textbf{Interface / metric} & \textbf{Frictional uptake} & \textbf{Indirect citation} \\
\midrule
BadukTV Phase 1 & -- & -- & -- & -- & -- \\
BadukTV Phase 2 & 0.38 & 76.39 & 23.61 & 0.00 & 1.39 \\
BadukTV Phase 3 & 1.35 & 44.54 & 59.06 & 1.09 & 0.98 \\
BadukTV Phase 4 & 2.63 & 53.67 & 49.86 & 2.62 & 1.33 \\
\midrule
Phase 4 institutional & 2.38 & 38.55 & 64.10 & 2.81 & 0.75 \\
Phase 4 creator field & 2.48 & 28.70 & 74.81 & 1.56 & 0.66 \\
\bottomrule
\end{tabular*}
\caption{Keyword-anchored AI-salient summary. The subset grows across phases, and the creator field leans further toward interface and metric mediation than institutional commentary.}
\label{tab:ai_salient_summary}
\end{table*}

\subsection{Institutional Longitudinal Backbone}
\label{sec:longitudinal}
The patterns below are anchored in \textit{BadukTV}, the only channel with consistent four-phase coverage. \textit{K-Baduk} and the creator channels enter only in Phases~3--4. We treat them as descriptive patterns within our stratified sample, not a census of the Go community, and the percentages are rates of overt AI activation, not estimates of total AI dependence.

\begin{figure}[t]
\centering
\includegraphics[width=\columnwidth]{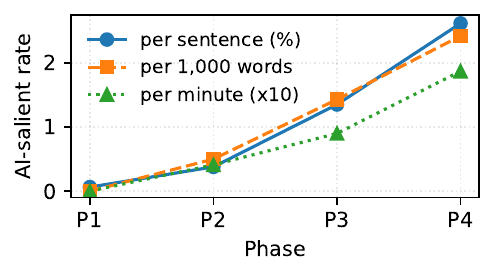}
\caption{BadukTV longitudinal AI-salient share across Phases~1--4. Per-minute is scaled $\times 10$ for comparability.}
\label{fig:longitudinal}
\end{figure}

The clearest longitudinal pattern is growth in the AI-salient subset itself. In \textit{BadukTV}, AI-salient sentences rise from an effectively zero pre-AlphaGo baseline to 0.38\% in Phase~2, 1.35\% in Phase~3, and 2.63\% in Phase~4 (Table~\ref{tab:ai_salient_summary}), suggesting that AI is increasingly verbally activated as a routine public explanatory resource rather than as an exceptional interruption. This pattern is robust to alternative periodization and unitization: under alternative Phase~3/4 boundaries at January~2022 and January~2024, \textit{BadukTV} still shows a clearly higher Phase~4 AI-salient share, and the strict subset rises from 0.50 to 1.43 to 2.43 per 1,000 words and from 0.041 to 0.090 to 0.188 per minute across Phases~2--4 (Figure~\ref{fig:longitudinal}).

\looseness=-1 The internal composition of that subset also changes (all figures below on $D_{Long}$). In Phase~2, the \textit{BadukTV} subset is cleanly divided into explicit-only (76.4\%) and interface-only (23.6\%) cases with no overlap ($N=72$). By Phase~3, interface-only (55.5\%) outnumbers explicit-only (40.9\%), with a small jointly anchored stratum (3.6\%; $N=916$). In Phase~4 \textit{BadukTV}, the two families become roughly comparable: 50.1\% explicit-only / 46.3\% interface-only / 3.5\% jointly anchored ($N=2{,}098$). Interface-only mediation thus peaks in Phase~3 (the diffusion period when KataGo overlays were newly visible), and Phase~4 sees a small rebound of explicit naming, concentrated in moments where AI is relayed as a quoted source or cited for a specific recommended move. The shift is sharper when \textit{K-Baduk} is pooled with \textit{BadukTV}: the Phase~4 institutional sample is interface-dominated at 35.9\% / 61.5\% / 2.6\% ($N=4{,}125$). Once AI is overtly verbally activated, commentary increasingly renders machine judgment through graphs, winrates, point-gap estimates, and other publicly legible cues.

Frictional uptake values in Table~\ref{tab:ai_salient_summary} are reported descriptively rather than as a basis for inferential claims; the Appendix~F diagnostic attributes most validation disagreement to Go-context-dependent sub-types. The more secure longitudinal result is simpler: AI becomes steadily more present in commentary, and that presence shifts from sparse novelty mentions toward a more normalized mixture of naming and interface rendering.

\subsection{Phase 4 Field-Level Contrast: Institutional and Creator-Led Mediation}
\label{sec:channel_comparison}

\looseness=-1 Under the keyword-anchored design, the Phase~4 contrast between institutional and creator-led commentary becomes clearer but also narrower. Per-sentence AI-salient share is very similar across the two settings: aggregated over Phase~4, institutional videos devote 2.38\% of all sentences to AI-salient talk, compared with 2.48\% in the broader creator field, and the video-level contrast is small relative to within-field variation ($t = -0.69$, $p = 0.493$). Under time- and word-based denominators, however, the creator field is higher: 2.97 versus 2.18 AI-salient sentences per 1,000 words ($t = 3.38$, $p < 0.001$) and 0.222 versus 0.176 per minute ($t = 2.39$, $p = 0.018$). Field-level volume is thus denominator-sensitive. Creator commentary is not more AI-saturated sentence for sentence, but it packs more AI-salient sentences into the same speaking time, consistent with faster solo narration rather than with a different rate of overt AI naming. We therefore place more weight on the compositional contrast than on any claim about volume.

Inside that subset, however, the two settings share dominant patterns and differ only at the margins. Both institutional and creator-led commentary are dominated by interface and metric mediation, with comparatively low rates of indirect citation. The overlap breakdown matters: in Phase~4, institutional subsets are 35.9\% explicit-only, 61.5\% interface-only, and 2.6\% jointly anchored, whereas the broader creator field is 25.2\% explicit-only, 71.3\% interface-only, and 3.5\% jointly anchored. The clearer field-level difference therefore lies in composition rather than volume: institutional subsets retain more explicit AI naming, whereas creator subsets lean more heavily on interface and metric rendering once AI becomes visible. This field-level picture survives alternative boundary shifts, and the main compositional tendencies do not reverse under a January~2024 cutoff.

\looseness=-1 To assess this contrast while honoring clustering, we model the binary outcome explicit-only(1) vs interface-only(0) at the sentence level (joint stratum excluded; $N=6{,}797$). We use a Generalized Estimating Equations (GEE) logistic regression with robust standard errors, which accounts for non-independence of sentences within videos (or videos within channels) that ordinary logistic regression would ignore. With cluster = video (172 clusters; exchangeable working correlation), the model yields OR$_{\text{creator vs.\ institutional}} = 0.66$, 95\% CI $[0.48, 0.91]$, $p = 0.010$: a creator-field AI-salient sentence has roughly a third lower odds of being explicit-only rather than interface-only. The same model with cluster = channel (7 clusters) gives an estimate in the same direction (OR $= 0.66$, $[0.26, 1.66]$) but is no longer significant ($p = 0.37$). With this channel count, no field-level claim about institutional vs.\ creator \emph{mediation norms} is supported by these data. The contrast we report generalizes to videos within our corpus rather than to the populations the channel labels imply. Per-video Welch tests on raw proportions agree (explicit-only 33.2\% institutional vs.\ 24.8\% creator, $t = -2.40$, $p = 0.016$; interface-only 64.6\% vs.\ 73.0\%, $t = 2.32$, $p = 0.020$). The composition contrast is therefore modest in size and statistically supported only at the video-clustered level. An aggregated $\chi^2$ test that ignores clustering produces $p < 10^{-16}$ and overstates the certainty (reported in Appendix~G). For AIES purposes, the modest-but-direction-consistent gap still matters because explicit naming and interface rendering make different public offers of transparency: one foregrounds the machine source, while the other folds it into familiar metrics and explanatory flow.

\begin{figure}[t]
\centering
\includegraphics[width=\columnwidth]{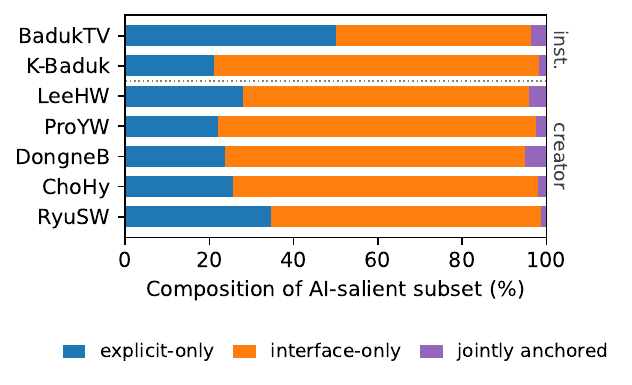}
\caption{Phase 4 channel-level composition of the AI-salient subset, by field. Channel codes: LeeHW = LeeHyunWookTV; ProYW = ProYeonwoo; RyuSW = RyuSihunWorld; ChoHy = ChoHyeyeon; DongneB = DongneBaduk.}
\label{fig:phase4_composition}
\end{figure}

This pattern also fits the channel-level heterogeneity (Figure~\ref{fig:phase4_composition}). In Phase~4, \textit{ProYeonwoo} and \textit{K-Baduk} are both strongly interface-heavy, but for different media reasons: the former often folds winrate and count talk directly into fast creator narration, whereas the latter dwells on graphs and interface cues within more explanatory television talk. \textit{LeeHyunWookTV}, \textit{RyuSihunWorld}, and \textit{ChoHyeyeon} retain relatively more explicit AI naming inside their subsets, but the broader creator field is not uniformly naming-heavy.

\looseness=-1 Decomposing the interface-only stratum into sub-types makes the field-level contrast sharper. Within Phase~4 interface-only AI-salient sentences ($N = 8{,}786$), creator-led commentary cites \emph{winrate} markers (e.g., \textit{승률}, percentage expressions) at 58.2\% versus 30.5\% in institutional commentary, and \emph{bluespot} (AI-recommended-move) markers at 18.4\% versus 11.2\%. Institutional commentary, by contrast, references the \emph{graph} itself (59.8\% vs.\ 21.1\% in the creator field). Per-sub-type GEE logistic regressions confirm the contrasts at the channel-clustered level despite only seven channels. Winrate (OR$_\text{creator vs.\ inst} = 4.66$, 95\% CI $[2.32, 9.35]$, $p < 10^{-4}$) and graph (OR $= 0.11$, $[0.04, 0.35]$, $p = 0.00013$) both survive Bonferroni correction at $\alpha = 0.05/4$. Bluespot is significant at the video-clustered level ($p = 0.003$) and nominally so at the channel level (OR $= 1.65$, $[1.04, 2.61]$, $p = 0.033$), but does not survive the same correction. Point-gap shows no field effect. The field-level contrast that was modest at the overall composition level is therefore substantial at the sub-type level, and the channel-clustered statistical caveat we noted above is partially relaxed for winrate and graph reference.

The compositional difference is descriptively meaningful. In creator-led channels, the interface-only share reaches 71.3\% (a practice in which AI recommendations and winrate overlays are displayed and narrated without explicit source attribution). We describe this pattern as \textit{source naturalization}: machine judgment enters public talk as the unmarked evaluative ground rather than as a quoted external source. Explicit naming, by contrast, preserves marked moments of source attribution. Whether such marked moments matter normatively depends on what is at stake: in Go's low-stakes viewing environment, where AI is also treated as a near-oracle ground truth, naturalization is the functional endpoint of domestication, not necessarily a failure of transparency. In higher-stakes domains where errors carry real-world harm and AI lacks Go's near-oracle status, the same compositional pattern raises governance concerns (a transfer we develop in the discussion). The format of mediation is, at minimum, a variable that travels with media context: institutional formats with multi-commentator cross-checking and editorial oversight retain more marked moments than creator formats organized around solo narration.

\subsection{Within-Speaker Follow-Up Test}
\label{sec:case_study}
\looseness=-1 The within-speaker comparison is best read as a follow-up probe rather than as a central result. To hold speaker identity as constant as possible, we compare Lee Hyunwook across two media settings (see Ethical Considerations on the rationale for using a single named commentator): appearing as a guest expert on \textit{BadukTV} versus hosting his own channel \textit{LeeHyunWookTV}. Using the paired corpus ($D_{Case}$), we analyze 48 temporal pairs with non-zero AI-salient discourse on both sides.

\looseness=-1 Under the stricter AI-salient measure, the same commentator looks more stable in \emph{volume} across settings than the earlier topic-family analysis suggested. Overall AI-salient share is very similar across the paired corpora (2.07\% personal versus 1.98\% television; paired $t = 0.45$, $p = 0.653$), and that stability holds under per-1,000-word and per-minute denominators ($p = 0.854$ and $p = 0.309$). \emph{Composition} moves modestly in the same direction as the field-level contrast (interface-only 70.3\% personal vs.\ 64.2\% television; paired $t = 2.21$, $p = 0.032$ on the 48 pairs with non-zero AI-salient discourse on both sides). Two caveats limit the strength of this claim: the test is marginal on 48 pairs, and the pair restriction introduces a selection effect we do not adjust for. We read it as suggestive evidence of a scope condition (medium shapes \emph{how} machine judgment is rendered more than \emph{how often} it is verbally named), rather than as a confirmation of the field-level pattern.

\subsection{Exploratory Audience Uptake}
\label{sec:audience_uptake}
An exploratory event-linked chat analysis (48 event--baseline pairs, 619 hand-coded messages) suggests that AI uptake in live chat is substantially more common around AI-salient commentary events (20.4\% of messages) than in matched baseline windows (5.8\%; sign test $p < 0.001$). Because the sample is small, non-representative, and relies on single-coder categorization, we report full results and limitations in Appendix~H and treat the audience layer as a preliminary extension rather than a backbone claim.

\looseness=-1 Qualitatively, viewers in the coded event windows engage the machine source directly rather than merely echoing the commentator. Deference is the dominant register (``The AI's judgment is the proof''), but challenge also appears (``So KataGo's reading really is weak''), and viewers sometimes introduce parallel AI tools of their own. Viewers also name the AI around interface-only commentary events, re-marking a source the commentator left unmarked. Because challenge cases are sparse, we read these exchanges as illustrations that overt AI activation leaves viewers a discursive anchor for both deference and challenge, not as evidence of how often that anchor is used.

\subsection{Micro-Patterns in AI-Salient Commentary}
\label{sec:micro_patterns}
The keyword backbone is intentionally coarse, so we also inspected the AI-salient subset qualitatively to identify recurrent discourse moves that the aggregate measures flatten. These patterns are not the paper's main quantitative backbone, but they reveal three recurrent mediation practices through which commentators make machine judgment public.

First, \textit{reportive relay} lets the machine speak through ordinary commentary. In the expanded-only audit, Korean reportive forms such as ``$\sim$라고 하네요'' and ``$\sim$라고 합니다'' frequently relayed machine-backed move recommendations without repeatedly naming the source. A \textit{BadukTV} commentator says, ``It says to block on this side now,'' and a creator-side commentator says, ``If this comes out, it says White should play here first before taking this,'' each closing on a reportive ending rather than a named source. These are not merely quotations. They are routinized relay forms that let machine judgment enter public talk through familiar reportive Korean.

Second, \textit{translational mediation} re-encodes machine outputs into viewer-legible game language. A creator commentator explains, ``Put simply, even if the point gap shrinks, Black's chances of winning rise,'' translating a graph movement into a more intelligible game-state interpretation. Here the commentator is not just reading off the interface, but actively converting machine-facing metrics into a public explanation.

Third, \textit{human--machine contrast} marks the boundary between what the AI recommends and what humans can realistically find or play. Commentators say, ``If you play this way it goes to 55\%, but a human cannot really play it that way,'' or ``It is the answer the AI gives, but honestly it is hard for a human to find.'' Such contrasts keep machine authority visible while simultaneously qualifying it through human playability and judgment.

\looseness=-1 These forms also sound different across phases. In Phase~2, when explicit naming dominated the subset (76.4\%), AI enters commentary as a topic to be appraised rather than as a source to be relayed. A \textit{BadukTV} commentator grants that ``AI clearly has weaknesses, but so far we have not been able to find them,'' and another, live during the first AlphaGo game, judges that ``it might take AI longer to catch up to humans than we thought.'' The machine is what the commentary is about, named because its judgment is still under negotiation. In Phase~4, the same channel relays machine judgment without naming its source (``It says to block on this side now''), and explicit naming survives mainly where a specific recommendation or a human--machine boundary is at stake. Across the decade, AI moves from what commentary is \emph{about} to what commentary speaks \emph{from}.

Beyond these overt patterns, much AI mediation travels through silent visual infrastructure that our keyword backbone does not capture. In a typical late-routine scene, the winrate graph is permanently displayed on screen and the commentator narrates a sharp graph movement, such as ``Oh, it widens a lot here,'' without naming AI at all. On creator channels, AI-recommended moves are often overlaid directly on the board, and the commentator simply says ``This side looks good'' while pointing at an AI-generated suggestion, making it impossible to distinguish personal judgment from machine recommendation. These scenes illustrate why the measured 2.63\% should not be read as a measure of AI dependence: what our strict backbone captures is the smaller set of moments when the visual AI layer rises into marked, overt speech.

\section{Discussion and Conclusion}

\subsection{Interpretive Synthesis}
\looseness=-1 Our evidence covers changes in overt verbal AI activation and in the mix of mediation forms; it does not reach legitimacy or governance outcomes. Within that scope, the composition of AI-facing talk is more informative than its volume: institutional and creator settings devote similar sentence-level shares to it but differ in the explicit-naming vs.\ interface-rendering mix, and late-routine mediation increasingly travels through reportive, translational, and contrastive forms rather than repeated explicit naming.

These patterns also reframe the small size of the strict subset. Read against the already-embedded visual AI layer, the longitudinal trajectory is not ``more AI citation'' but a shift from marked novelty toward unmarked infrastructure, with late-routine naming concentrated around relay and recommendation rather than continuous narration.

\looseness=-1 We read this trajectory as the \emph{communicative signature} of domestication, not as a direct test of its causal substrate. In modern broadcasting, screen and voice are co-constitutive, and avoiding verbal redundancy with an on-screen graphic is a standard media logic. The pattern we observe could in principle reflect a deeper socio-cognitive change in how experts treat AI judgment, or a more local adaptation to broadcast-UI conventions. Our verbal-only data describe the shape of the public mediation layer rather than discriminate between these two readings.

\subsection{A Mediation-Form Typology}
The qualitative micro-patterns and compositional analysis point toward a structured typology of public AI mediation. Table~\ref{tab:mediation_typology} organizes the five recurrent forms along two axes: whether they foreground or recede the machine source, and whether they anchor meaning in metric or narrative semiotic modes.

\begin{table}[t]
\centering
\footnotesize
\setlength{\tabcolsep}{3pt}%
\begin{tabular*}{\columnwidth}{@{\extracolsep{\fill}}lp{3.15cm}p{3.15cm}@{}}
\toprule
 & \textbf{Source-foregrounding} & \textbf{Source-receding} \\
\midrule
\textbf{Metric} & Translation (re-encodes metric output for viewers) & Interface rendering (graphs, winrates naturalized into flow) \\
\addlinespace
\textbf{Narrative} & Naming; Human--machine contrast (marks source or boundary) & Reportive relay (machine judgment enters via ordinary grammar) \\
\bottomrule
\end{tabular*}
\caption{Our mediation-form typology.}
\label{tab:mediation_typology}
\end{table}

\looseness=-1 Each form has a distinct affordance: \textit{naming} foregrounds attribution; \textit{interface rendering} maximizes usability but risks \textit{naturalization}; \textit{reportive relay} routinizes machine authority through ordinary speech patterns; \textit{translation} foregrounds the human interpreter while re-encoding metrics; and \textit{human--machine contrast} preserves contestability by marking the AI/human boundary. A limiting case is \textit{strategic non-mention}: following an AI recommendation without naming it at all, invisible to our keyword backbone and beyond our verbal evidence. Source-foregrounding forms preserve what we call \textit{hooks of contestability}: discursive anchors through which audiences can recognize, question, or challenge the algorithmic source. Source-receding forms increase fluency at the cost of leaving fewer marked moments for challenge. When explicit naming appears alongside surprise or uncertainty markers, the AI name itself anchors the contested move. Once the same move is relayed through ``$\sim$라고 합니다'', that anchor disappears from the commentary itself. The longitudinal shift toward more interface-heavy mediation within \textit{BadukTV} (\S\ref{sec:longitudinal}) is, descriptively, the communicative trace of domestication. Whether it carries governance weight depends on the domain into which the same forms are transported.

\subsection{Governance Implications}
\looseness=-1 A careful reading of our findings requires separating descriptive, functional, and normative claims. Descriptively, late-routine Korean Go commentary shows a shift from source-foregrounding toward source-receding mediation, naturalization in our terms. Functionally, this is consistent with successful domestication \cite{silverstone1996design,berker2006domestication,hartmann2013domestication}: in a complete-information game with low real-world stakes, ubiquitous on-screen interfaces, and an audience that already assumes machine evaluation stands behind any authoritative judgment, the recession of explicit naming reflects an equilibrium in which the machine source no longer needs to be marked. The normative question (whether this constitutes opacity, erosion of contestability, or appropriate fluency) is therefore not settled by the Go evidence alone. It depends on whether the same mediation forms travel into settings where the conditions that make naturalization safe in Go no longer hold.

\looseness=-1 This typology reframes explainability for AIES in two ways. First, it suggests that explanation should be studied not only as information attached to a system, but as a \textit{situated interpretive activity performed by intermediaries} \cite{ehsan2021expanding,kaur2020interpreting,menon2024lessons}. In our data, commentators do not merely transmit AI outputs: they relay, translate, and qualify them for viewers. That is a form of \textit{delegated interpretation}: human experts become the public-facing layer through which model outputs acquire meaning and practical force.

\looseness=-1 Second, our evidence is from a single domain, but we close with brief reflections on how the typology might travel. What the Go case establishes is descriptive: a repertoire of public mediation forms and a documented shift in their mix under near-oracle, low-stakes conditions. What the typology offers beyond Go is a set of predictions about which forms should dominate as reliability and stakes change, and those predictions are not tested here. Go is unusually clean for an AI deployment setting: as a complete-information game with an objective ground truth that AI is known to evaluate near-optimally, it makes interface rendering a relatively safe form, because the underlying source is itself unusually trustworthy. Where domains depart from that condition, we would expect the weight of mediation forms to shift: toward translation and naming in medicine, where the mediator must manage epistemic uncertainty alongside the score \cite{cai2019hello,lebovitz2022engage}; toward naming under disclosure obligation in law and risk assessment \cite{green2019disparate,angwin2016machine}; and toward interface rendering in finance, with naturalization risk where ``the algorithm estimates your risk at'' collapses into ``your risk score'' \cite{stark2019data}. The pattern we observe rests on a specific empirical condition: AI's evaluations in Go are demonstrably near-optimal, and the community has lived with that demonstration for nearly a decade. Whether the same forms remain functional where comparable reliability is not yet established is for future work to settle \cite{almada2019human,lyons2021conceptualising,raji2020closing}.

\subsection{Media Context}
\looseness=-1 Field-level results suggest that media context shapes the \textit{composition} of AI mediation more than its \textit{volume}: the institutional/creator split between explicit naming and interface rendering tracks media logic (multi-commentator editorial formats vs.\ solo creator narration). The within-speaker probe points in the same direction, but as suggestive evidence of a scope condition rather than confirmation of the field-level pattern.

\looseness=-1 The sub-type contrasts (\S\ref{sec:channel_comparison}) suggest a division of mediation labor: institutional commentary anchors on the graph itself, narrating its shifts (e.g., asking ``What's happening with the graph...?'') and developing analytic explanation around that anchor, while creator-led commentary reads off the AI-derived numbers directly, delegating more of the analytic content to the AI output itself.

\subsection{Limitations}
\looseness=-1 The corpus is Korean-language only and domain-specific. The longitudinal backbone is \textit{BadukTV}-centered and captures high-visibility public commentary rather than the full Go-video population. The phases are unequal in span: Phases~1 and~2 cover months of 2016, while Phase~4 runs from 2021 to 2025. Main patterns survive alternative Phase~3/4 boundaries from January~2020 to January~2024, and restricting the Phase~4 institutional sample to the creator field's 2023--2024 window leaves its AI-salient share essentially unchanged (both in Appendix~D), so the pooling does not appear to drive the Phase~4 figures. The absolute volume trajectory nonetheless remains partially co-determined by production-side changes we do not measure.

\looseness=-1 Methodologically, the keyword backbone prioritizes precision and likely undercounts AI-relevant commentary lacking overt lexical markers. A multimodal analysis would reveal a larger AI-mediated layer. Relatedly, the institutional/creator sub-type contrast (\S\ref{sec:channel_comparison}) is observed only verbally. We cannot rule out that broadcast production (e.g., bluespot visibility) differs across channel types in ways that would mechanically produce part of the verbal differences we observe; disentangling the two readings requires reliable channel-level visual-element data. Whisper Base, our STT model, is the smallest open variant, and larger variants would transcribe Korean more accurately. Our keyword-side audits bound captured-set precision but not corpus-level recall, and AI mention by speakers with strong accents or in fast creator narration may be undercounted.

Primary coding was single-coder. The second-coder reliability results in Methods bound this risk, and we treat frictional uptake as a descriptive marker only. Some claims about routine AI use in professional Go rest on industry documentary and trade-press sources rather than peer-reviewed studies. The audience extension is preliminary.

\subsection{Conclusion}
\looseness=-1 A decade of Korean Go commentary on YouTube shows what happens after AI surpasses elite human performance and settles into routine infrastructure. AI-salient talk accounts for only $2.63\%$ of \textit{BadukTV} Phase~4 sentences while winrate graphs are visible for $97.99\%$ of broadcast time. The source label recedes while the metric talk persists, and we read that recession as the communicative signature of domestication. The mediation-form typology distinguishes source-foregrounding from source-receding practices and frames the conditions (low stakes, near-oracle AI, and machine involvement the audience already assumes) that make naturalization safe in Go. Where those conditions weaken, the same compositional pattern raises governance concerns about how machine judgment is rendered publicly intelligible, attributable, and contestable.

\section*{Ethical Considerations Statement}

All data are public observational trace data. No intervention or recruitment was involved. Usernames are omitted, chat is reported in aggregate, and quoted material is limited to short analytic illustrations. Viewer chat messages are quoted in English translation only, without usernames, to limit re-identification of individual viewers. We release code, keyword rules, per-video metadata, and validation templates, but not raw media, transcripts, chat text, or user identifiers due to platform, privacy, and copyright constraints.

All speech we analyze is drawn from publicly accessible YouTube content produced for broad audiences: commercial broadcasts on cable Go channels and public-facing creator-led channels. The ethical framing is therefore closer to analysis of a public figure's broadcast speech than to study of private communication. One section further examines a single named commentator (Lee Hyunwook) across two of his public broadcasting contexts. This within-speaker analysis was unavoidable because he is the only commentator in our sample with sustained coverage across both an institutional and a creator-led setting. We have nonetheless adjusted analytic and writing scope to minimize avoidable harm: we restrict ourselves to aggregate descriptive patterns rather than evaluative judgments about individual style, quote only short illustrative fragments of broadcast speech, do not attribute private context to any commentator, and make no normative prescriptions about how any individual ought to speak.

\section*{Researcher Positionality Statement}

I am a Korean native speaker with a non-professional but extended relationship to Go: a few years of children's Go academy in elementary school, occasional play with adults afterwards, a long lapse through middle school, and a return to the game through the manga \emph{Hikaru no Go}, followed by years of casual online play and watching competitive tournament coverage. I am therefore an enthusiast rather than a professional. I am linguistically and culturally inside the data, but only partially overlapping with the professional commentators I analyze. I experienced the AlphaGo--Lee Sedol match as a personal worldview break. I had actually played the game and shared the common-sense view that intuitive positional judgment was the bottleneck AI would not soon cross, so the loss made questions about AI capability newly serious for me. The paper's marked-form / domestication framing, however, did not emerge from the loss itself. It crystallized gradually over the years that followed, as I watched how the Korean Go world (its commentary, its YouTube channels, its routinized AI references) adapted to AI becoming infrastructure. I report it as an interpretive lens rather than as a neutral derivation.

Methodologically, I come from computational social science and the data-driven study of online social media. This paper applies that toolkit to a domain (Korean Go on YouTube) I happen to know linguistically and culturally. I have not tried to set these views aside, because I do not think it is honest to claim I could. The second-coder reliability check (Methods) is the empirical counterweight I report against this interpretive position: a Korean speaker who does not follow Go reproduced the coding without sharing my history with the game.

Reflexively, I am working through an analogous transition myself. As a data-driven researcher I encounter AI coding assistants with the same mixture of unease and curiosity that the Korean Go community lived through a decade earlier, facing AI systems that materially surpass the intuition their expertise is built on. For them it arrived at once; for me it is arriving slowly. The trust-building, slow contestation, and gradual renegotiation of expert authority that Korean Go professionals have already worked out is, from my position, lived expertise from a transition I am myself still inside. This study is therefore best read not as a diagnosis of how a community ``coped'' with AI, but as careful listening to a community that has already traveled further along a road many professional fields, including my own, are only beginning to walk.

\section*{Adverse Impact Statement}

The mediation-form typology developed here could, in principle, be used adversarially: a vendor or platform optimizing AI deployment for fluency and uptake has every incentive to push commentary, advice, or interface design toward the source-receding side of our typology (interface rendering, reportive relay, strategic non-mention). We flag this risk explicitly because the same analytic vocabulary is useful for audit and for optimization, and only the auditing use preserves \emph{hooks of contestability}. Our intended use is the former. We do not provide design recommendations for the latter.

We are aware that visible patterns identified here can travel back into broadcasting norms or platform policy in ways that retroactively pressure commentators' practice. We do not endorse such pressures, and the descriptive composition we observe is not a prescription.

Finally, the \emph{domestication} framing is doubly available. It supports a critical reading in which the recession of marked AI naming erodes contestability in higher-stakes domains, and it equally supports a vendor-friendly reading in which AI fluency is presented as a sign of mature integration. Both inferences can be drawn from the same data. The cross-domain reflections above are offered as questions for empirical follow-up, not as design prescriptions, and we ask future work to make the audit-vs-optimization distinction explicit when borrowing the vocabulary.

\section*{Acknowledgments}

This research was supported by the MSIT (Ministry of Science, ICT), Korea, under the Global Research Support Program in the Digital Field program (RS-2024-00425354) supervised by the IITP (Institute for Information \& Communications Technology Planning \& Evaluation).

\bibliography{aaai2026}

\begin{thebibliography}{50}
\providecommand{\natexlab}[1]{#1}

\bibitem[{Alfrink et~al.(2023)Alfrink, Keller, Kortuem, and
  Doorn}]{alfrink2023contestable}
Alfrink, K.; Keller, I.; Kortuem, G.; and Doorn, N. 2023.
\newblock Contestable AI by Design: Towards a Framework.
\newblock \emph{Minds and Machines}, 33: 613--639.

\bibitem[{Almada(2019)}]{almada2019human}
Almada, M. 2019.
\newblock Human Intervention in Automated Decision-Making: Toward the
  Construction of Contestable Systems.
\newblock In \emph{Proceedings of the Seventeenth International Conference on
  Artificial Intelligence and Law}, 2--11. ACM.

\bibitem[{Angwin et~al.(2022)Angwin, Larson, Mattu, and
  Kirchner}]{angwin2016machine}
Angwin, J.; Larson, J.; Mattu, S.; and Kirchner, L. 2022.
\newblock Machine bias.
\newblock In \emph{Ethics of data and analytics}, 254--264. Auerbach
  Publications.

\bibitem[{{BadukTV}(2026)}]{baduktvchannel}
{BadukTV}. 2026.
\newblock BadukTV YouTube Channel.
\newblock \url{https://www.youtube.com/@baduk_tv}.
\newblock Accessed March 29, 2026.

\bibitem[{Berker et~al.(2006)Berker, Hartmann, Punie, and
  Ward}]{berker2006domestication}
Berker, T.; Hartmann, M.; Punie, Y.; and Ward, K.~J., eds. 2006.
\newblock \emph{Domestication of Media and Technology}.
\newblock Maidenhead: Open University Press.

\bibitem[{Brauner et~al.(2025)Brauner, Glawe, Liehner, Vervier, and
  Ziefle}]{brauner2025mapping}
Brauner, P.; Glawe, F.; Liehner, G.~L.; Vervier, L.; and Ziefle, M. 2025.
\newblock Mapping public perception of artificial intelligence: Expectations,
  risk--benefit tradeoffs, and value as determinants for societal acceptance.
\newblock \emph{Technological Forecasting and Social Change}, 220: 124304.
\newblock Doi:10.1016/j.techfore.2025.124304.

\bibitem[{Cai et~al.(2019)Cai, Winter, Steiner, Wilcox, and
  Terry}]{cai2019hello}
Cai, C.~J.; Winter, S.; Steiner, D.; Wilcox, L.; and Terry, M. 2019.
\newblock ``Hello AI'': Uncovering the Onboarding Needs of Medical
  Practitioners for Human-AI Collaborative Decision-Making.
\newblock \emph{Proceedings of the ACM on Human-Computer Interaction}, 3(CSCW):
  1--24.

\bibitem[{Castleman and Korolova(2024)}]{castleman2024why}
Castleman, J.; and Korolova, A. 2024.
\newblock Why Am {I} Still Seeing This: Measuring the Effectiveness of Ad
  Controls and Explanations in {AI}-Mediated Ad Targeting Systems.
\newblock In \emph{Proceedings of the AAAI/ACM Conference on AI, Ethics, and
  Society}, volume~7, 255--266. AAAI Press.

\bibitem[{{ChoHyeyeon}(2026)}]{chohyeyeonchannel}
{ChoHyeyeon}. 2026.
\newblock Cho Hyeyeon Pro 9-dan YouTube Channel.
\newblock \url{https://www.youtube.com/channel/UCaVMWUQcMRqNxNrsBhhS85A}.
\newblock Accessed July 30, 2026.

\bibitem[{Dietvorst, Simmons, and Massey(2015)}]{dietvorst2015algorithm}
Dietvorst, B.~J.; Simmons, J.~P.; and Massey, C. 2015.
\newblock Algorithm aversion: people erroneously avoid algorithms after seeing
  them err.
\newblock \emph{Journal of experimental psychology: General}, 144(1): 114.

\bibitem[{{DongneBaduk}(2026)}]{dongnebadukchannel}
{DongneBaduk}. 2026.
\newblock DongneBaduk YouTube Channel.
\newblock \url{https://www.youtube.com/@%EB%8F%99%EB%84%A4%EB%B0%94%EB%91%91}.
\newblock Accessed July 30, 2026.

\bibitem[{Doshi-Velez and Kim(2017)}]{doshi2017towards}
Doshi-Velez, F.; and Kim, B. 2017.
\newblock Towards a rigorous science of interpretable machine learning.
\newblock \emph{arXiv preprint arXiv:1702.08608}.

\bibitem[{{EBS Documentary}(2026)}]{ebs2026badukdoc}
{EBS Documentary}. 2026.
\newblock 바둑계는 끝났다, 신의 경지까지 올라간 인공지능에
  절대 못 이긴다. 알파고 대국 10년 후 | 다큐프라임 |
  \#골라듄다큐.
\newblock \url{https://www.youtube.com/watch?v=gz2Ig7Jf88I}.
\newblock Uploaded March 11, 2026. Accessed April 18, 2026.

\bibitem[{Ehsan et~al.(2021)Ehsan, Liao, Muller, Riedl, and
  Weisz}]{ehsan2021expanding}
Ehsan, U.; Liao, Q.~V.; Muller, M.; Riedl, M.~O.; and Weisz, J. 2021.
\newblock Expanding Explainability: Towards Social Transparency in AI Systems.
\newblock In \emph{Proceedings of the 2021 CHI Conference on Human Factors in
  Computing Systems}, 1--19. ACM.

\bibitem[{Ensmenger(2012)}]{ensmenger2012chess}
Ensmenger, N. 2012.
\newblock Is chess the drosophila of artificial intelligence? A social history
  of an algorithm.
\newblock \emph{Social Studies of Science}, 42(1): 5--30.

\bibitem[{Epstein et~al.(2020)Epstein, Levine, Rand, and
  Rahwan}]{epstein2020gets}
Epstein, Z.; Levine, S.; Rand, D.~G.; and Rahwan, I. 2020.
\newblock Who gets credit for AI-generated art?
\newblock \emph{iScience}, 23(9).

\bibitem[{Gelly et~al.(2012)Gelly, Kocsis, Schoenauer, Sebag, Silver,
  Szepesv{\'a}ri, and Teytaud}]{gelly2012grand}
Gelly, S.; Kocsis, L.; Schoenauer, M.; Sebag, M.; Silver, D.; Szepesv{\'a}ri,
  C.; and Teytaud, O. 2012.
\newblock The grand challenge of computer Go: Monte Carlo tree search and
  extensions.
\newblock \emph{Communications of the ACM}, 55(3): 106--113.

\bibitem[{Gibson(1979)}]{gibson1979ecological}
Gibson, J.~J. 1979.
\newblock \emph{The ecological approach to visual perception}.
\newblock Houghton Mifflin.

\bibitem[{Glikson and Woolley(2020)}]{glikson2020human}
Glikson, E.; and Woolley, A.~W. 2020.
\newblock Human trust in artificial intelligence: Review of empirical research.
\newblock \emph{Academy of management annals}, 14(2): 627--660.

\bibitem[{Green and Chen(2019)}]{green2019disparate}
Green, B.; and Chen, Y. 2019.
\newblock Disparate Interactions: An Algorithm-in-the-Loop Analysis of Fairness
  in Risk Assessments.
\newblock In \emph{Proceedings of the Conference on Fairness, Accountability,
  and Transparency}, 90--99.

\bibitem[{Hartmann(2013)}]{hartmann2013domestication}
Hartmann, M. 2013.
\newblock From domestication to mediated mobilism.
\newblock \emph{Mobile Media \& Communication}, 1(1): 42--49.

\bibitem[{Hjarvard(2008)}]{hjarvard2008mediatization}
Hjarvard, S. 2008.
\newblock The Mediatization of Society: A Theory of the Media as Agents of
  Social and Cultural Change.
\newblock \emph{Nordicom Review}, 29(2): 105--134.

\bibitem[{Hutchby(2001)}]{hutchby2001technologies}
Hutchby, I. 2001.
\newblock Technologies, texts and affordances.
\newblock \emph{Sociology}, 35(2): 441--456.

\bibitem[{{K-Baduk}(2026)}]{kbadukchannel}
{K-Baduk}. 2026.
\newblock K-Baduk YouTube Channel.
\newblock \url{https://www.youtube.com/@kbaduktv}.
\newblock Accessed March 29, 2026.

\bibitem[{Kaur et~al.(2020)Kaur, Nori, Jenkins, Caruana, Wallach, and
  Vaughan}]{kaur2020interpreting}
Kaur, H.; Nori, H.; Jenkins, S.; Caruana, R.; Wallach, H.; and Vaughan, J.~W.
  2020.
\newblock Interpreting Interpretability: Understanding Data Scientists' Use of
  Interpretability Tools for Machine Learning.
\newblock In \emph{Proceedings of the 2020 CHI Conference on Human Factors in
  Computing Systems}, 1--14. ACM.

\bibitem[{{Korea Baduk Association}(2020)}]{baduktv2020format}
{Korea Baduk Association}. 2020.
\newblock 10대 핫이슈로 돌아본 2020년 바둑TV!
\newblock \url{https://m.baduk.or.kr/news/B02_view.asp?news_sum_no=7538}.
\newblock BadukTV news feature on 2020 broadcast-screen changes, including AI
  position evaluation standardization around KataGo. Published October 27,
  2020. Accessed April 18, 2026.

\bibitem[{{Korea Baduk Association}(2021)}]{monthlybaduk2021ai}
{Korea Baduk Association}. 2021.
\newblock 핫피플 | 한국바둑AI연구소 이현호 대표.
\newblock \url{https://m.baduk.or.kr/news/B03_view.asp?news_no=155}.
\newblock 월간바둑, published July 2, 2021. Accessed April 18, 2026.

\bibitem[{{Korea Baduk Association}(2024)}]{kim2024opening70}
{Korea Baduk Association}. 2024.
\newblock 김만수, 초반 공부 끝판왕 ‘초반 70수’ 출간.
\newblock \url{https://m.baduk.or.kr/news/B01_view.asp?news_no=4940}.
\newblock Published April 18, 2024. Accessed April 18, 2026.

\bibitem[{Lebovitz, Lifshitz-Assaf, and Levina(2022)}]{lebovitz2022engage}
Lebovitz, S.; Lifshitz-Assaf, H.; and Levina, N. 2022.
\newblock To engage or not to engage with AI for critical judgments: How
  professionals deal with opacity when using AI for medical diagnosis.
\newblock \emph{Organization science}, 33(1): 126--148.

\bibitem[{Lee(2024)}]{bab2min2024kiwi}
Lee, M.-c. 2024.
\newblock Kiwi: Developing a Korean morphological analyzer based on statistical
  language models and skip-bigram.
\newblock \emph{Korean Journal of Digital Humanities}, 1(1): 109--136.

\bibitem[{{LeeHyunWookTV}(2026)}]{leehyunwooktvchannel}
{LeeHyunWookTV}. 2026.
\newblock LeeHyunWookTV YouTube Channel.
\newblock \url{https://www.youtube.com/@leehyunwooktv}.
\newblock Accessed March 29, 2026.

\bibitem[{Liao, Gruen, and Miller(2020)}]{liao2020questioning}
Liao, Q.~V.; Gruen, D.; and Miller, S. 2020.
\newblock Questioning the AI: Informing Design Practices for Explainable AI
  User Experiences.
\newblock In \emph{Proceedings of the 2020 CHI Conference on Human Factors in
  Computing Systems}, 1--15. ACM.

\bibitem[{Logg, Minson, and Moore(2019)}]{logg2019algorithm}
Logg, J.~M.; Minson, J.~A.; and Moore, D.~A. 2019.
\newblock Algorithm appreciation: People prefer algorithmic to human judgment.
\newblock \emph{Organizational Behavior and Human Decision Processes}, 151:
  90--103.

\bibitem[{Lyons, Velloso, and Miller(2021)}]{lyons2021conceptualising}
Lyons, H.; Velloso, E.; and Miller, T. 2021.
\newblock Conceptualising Contestability: Perspectives on Contesting
  Algorithmic Decisions.
\newblock \emph{Proceedings of the ACM on Human-Computer Interaction},
  5(CSCW1): 1--22.

\bibitem[{Menon et~al.(2024)Menon, Omar, Nahar, Papademetris, Fiellin, and
  K{\"a}stner}]{menon2024lessons}
Menon, A.~V.; Omar, Z.~A.; Nahar, N.; Papademetris, X.; Fiellin, L.~E.; and
  K{\"a}stner, C. 2024.
\newblock Lessons from Clinical Communications for Explainable {AI}.
\newblock In \emph{Proceedings of the AAAI/ACM Conference on AI, Ethics, and
  Society}, volume~7, 958--970. AAAI Press.

\bibitem[{Metcalf et~al.(2021)Metcalf, Moss, Watkins, Singh, and
  Elish}]{metcalf2021impact}
Metcalf, J.; Moss, E.; Watkins, E.~A.; Singh, R.; and Elish, M.~C. 2021.
\newblock Algorithmic Impact Assessments and Accountability: The
  Co-construction of Impacts.
\newblock In \emph{Proceedings of the 2021 ACM Conference on Fairness,
  Accountability, and Transparency}, 735--746. ACM.

\bibitem[{Miller(2019)}]{miller2019explanation}
Miller, T. 2019.
\newblock Explanation in artificial intelligence: Insights from the social
  sciences.
\newblock \emph{Artificial Intelligence}, 267: 1--38.

\bibitem[{Miyazaki et~al.(2024)Miyazaki, Murayama, Uchiba, An, and
  Kwak}]{miyazaki2024public}
Miyazaki, K.; Murayama, T.; Uchiba, T.; An, J.; and Kwak, H. 2024.
\newblock Public perception of generative AI on Twitter: an empirical study
  based on occupation and usage.
\newblock \emph{EPJ Data Science}, 13(1): 2.

\bibitem[{{ProYeonwoo}(2026)}]{proyeonwoochannel}
{ProYeonwoo}. 2026.
\newblock ProYeonwoo YouTube Channel.
\newblock \url{https://www.youtube.com/@proyeonwoo}.
\newblock Accessed March 29, 2026.

\bibitem[{Radford et~al.(2023)Radford, Kim, Xu, Brockman, McLeavey, and
  Sutskever}]{radford2023whisper}
Radford, A.; Kim, J.~W.; Xu, T.; Brockman, G.; McLeavey, C.; and Sutskever, I.
  2023.
\newblock Robust speech recognition via large-scale weak supervision.
\newblock In \emph{International conference on machine learning}, 28492--28518.
  PMLR.

\bibitem[{Rahwan(2018)}]{rahwan2018society}
Rahwan, I. 2018.
\newblock Society-in-the-Loop: Programming the Algorithmic Social Contract.
\newblock \emph{Ethics and Information Technology}, 20(1): 5--14.

\bibitem[{Raji et~al.(2020)Raji, Smart, White, Mitchell, Gebru, Hutchinson,
  Smith-Loud, Theron, and Barnes}]{raji2020closing}
Raji, I.~D.; Smart, A.; White, R.~N.; Mitchell, M.; Gebru, T.; Hutchinson, B.;
  Smith-Loud, J.; Theron, D.; and Barnes, P. 2020.
\newblock Closing the AI accountability gap: Defining an end-to-end framework
  for internal algorithmic auditing.
\newblock In \emph{Proceedings of the 2020 conference on fairness,
  accountability, and transparency}, 33--44.

\bibitem[{{RyuSihunWorld}(2026)}]{ryusihunworldchannel}
{RyuSihunWorld}. 2026.
\newblock Ryu Shikun's Go World YouTube Channel.
\newblock \url{https://www.youtube.com/channel/UCqcHLpDSKKSkQDg1xO0B4EA}.
\newblock Accessed July 30, 2026.

\bibitem[{Shneiderman(2020)}]{shneiderman2020hcai}
Shneiderman, B. 2020.
\newblock Human-Centered Artificial Intelligence: Reliable, Safe \&
  Trustworthy.
\newblock \emph{International Journal of Human--Computer Interaction}, 36(6):
  495--504.

\bibitem[{Shoemaker and Vos(2009)}]{shoemaker2009gatekeeping}
Shoemaker, P.~J.; and Vos, T.~P. 2009.
\newblock \emph{Gatekeeping Theory}.
\newblock New York: Routledge.

\bibitem[{Silver et~al.(2016)Silver, Huang, Maddison, Guez, Sifre, Van
  Den~Driessche, Schrittwieser, Antonoglou, Panneershelvam, Lanctot
  et~al.}]{silver2016mastering}
Silver, D.; Huang, A.; Maddison, C.~J.; Guez, A.; Sifre, L.; Van Den~Driessche,
  G.; Schrittwieser, J.; Antonoglou, I.; Panneershelvam, V.; Lanctot, M.;
  et~al. 2016.
\newblock Mastering the game of Go with deep neural networks and tree search.
\newblock \emph{nature}, 529(7587): 484--489.

\bibitem[{Silverstone and Haddon(1996)}]{silverstone1996design}
Silverstone, R.; and Haddon, L. 1996.
\newblock Design and the domestication of information and communication
  technologies: Technical change and everyday life.
\newblock In \emph{Communication by design}, 44--74. Oxford University Press.

\bibitem[{Skitka, Mosier, and Burdick(1999)}]{skitka1999does}
Skitka, L.~J.; Mosier, K.~L.; and Burdick, M. 1999.
\newblock Does automation bias decision-making?
\newblock \emph{International Journal of Human-Computer Studies}, 51(5):
  991--1006.

\bibitem[{Stark and Hoffmann(2019)}]{stark2019data}
Stark, L.; and Hoffmann, A.~L. 2019.
\newblock Data is the New What? Popular Metaphors and Professional Ethics in
  Emerging Data Culture.
\newblock \emph{Journal of Cultural Analytics}, 4(1).

\bibitem[{Wu(2019)}]{wu2019accelerating}
Wu, D.~J. 2019.
\newblock Accelerating self-play learning in Go.
\newblock \emph{arXiv preprint arXiv:1902.10565}.

\end{thebibliography}

% The supplement is typeset as its own document (single column, 1in margins).
% It is appended here so the arXiv version carries the Appendices A-I that
% the main text refers to. Generated by make_arxiv_package.sh; do not edit.
\includepdf[pages=-]{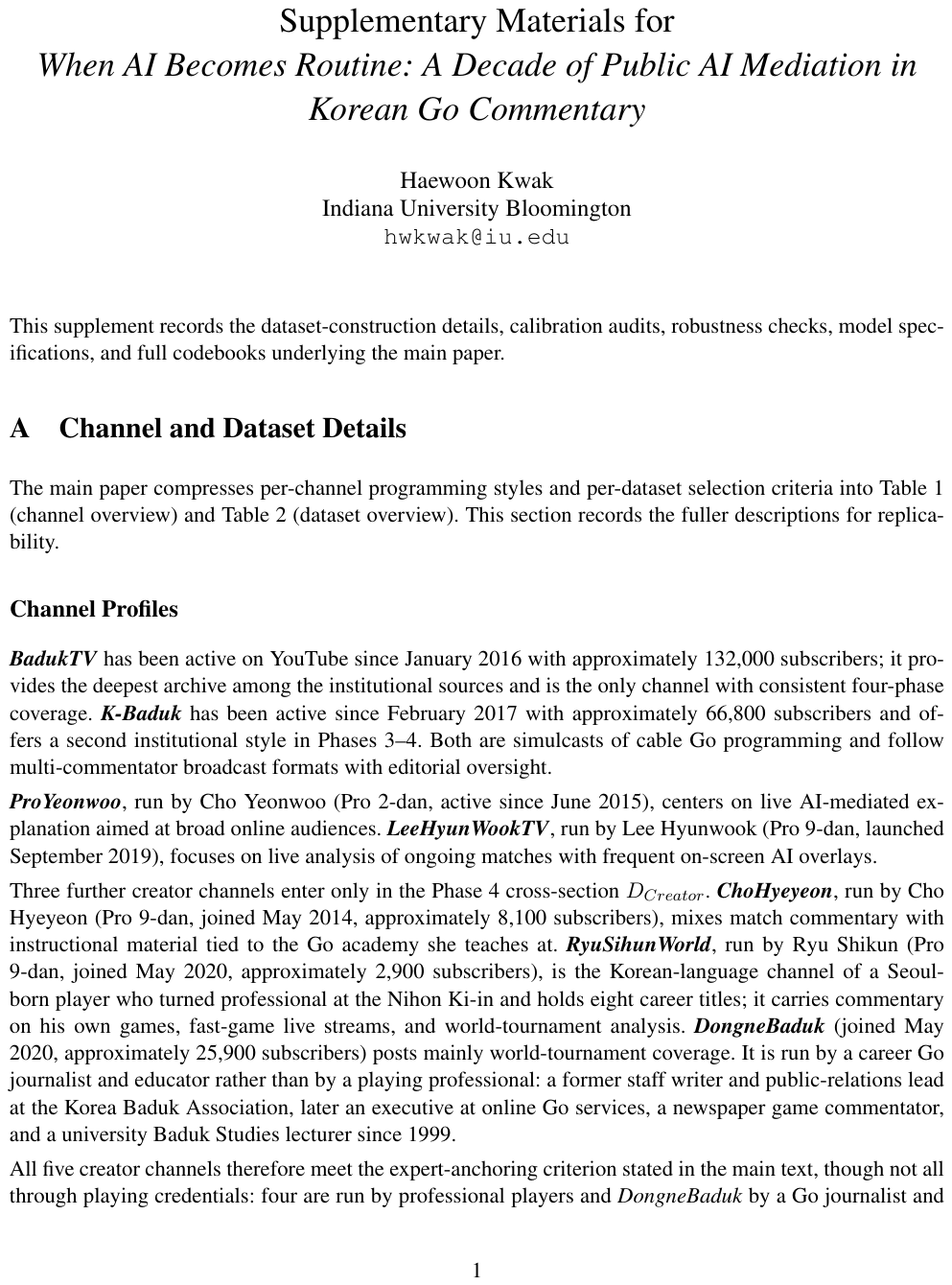}

\end{document}